\documentclass[twoside,11pt] {article}

\usepackage{graphics}
\usepackage{amssymb}
\usepackage{amsmath}

\begin{document}

\title{On coordinate distances to type Ia supernovae\\ and radio galaxies}
\author{V.E. Kuzmichev and V.V. Kuzmichev\\[0.5cm]
\itshape Bogolyubov Institute for Theoretical Physics,\\
\itshape Nat. Acad. of Sci. of Ukraine, Kiev, 03143 Ukraine}
\date{}

\maketitle

\pagestyle{myheadings} \thispagestyle{plain} \markboth{V.E.
Kuzmichev and V.V. Kuzmichev}{On coordinate distances}
\setcounter{page}{1}

\begin{abstract}
The quantum model of the homogeneous and isotropic universe
predicts logarithmic-law dependence of coordinate distance to
source on redshift $z$ which is in good agreement with type Ia
supernovae and radio galaxies observations for the redshift range
$z = 0.01 - 1.8$. A comparison with phenomenological models with
dark energy in the form of cosmological constant and without dark
energy component is made. Fluctuations of the cosmological scale
factor about its average value which can arise in the early
universe produce accelerating or decelerating expansions of space
subdomains containing separate sources with high redshift whereas
the universe as a whole expands at a steady rate.
\end{abstract}


\section{Introduction}
\label{Intro}

The high-redshift type Ia supernova (SN Ia) observations
\cite{Rie,Per,Ton} can be explained by an accelerating expansion
of the present-day universe \cite{Rie,Per,Tu1,Pee,Liv}. Such
conclusion assumes that observed dimming of the SNe Ia is hardly
caused by physical phenomena non-related to overall expansion of
the universe as a whole (see e.g. Refs. \cite{Ton,Vish} for
discussion and review). Furthermore it is supposed that matter
component of energy density in the universe $\rho_{M}$ varies with
the expansion of the universe as $a^{-3}$ (i.e. it has practically
vanishing pressure, $p_{M} \approx 0$), where $a$ is the
cosmological scale factor, while mysterious cosmic fluid
(so-called dark energy \cite{Tu2,Ost}) is describes by the
equation of state $p_{X} = w_{X} \rho_{X}$, where $-1 \leq w_{X}
\leq - \frac{1}{3}$ \cite{Tu1,Pee}. In the models with the
cosmological constant ($\Lambda$CDM) one has $w_{X} = - 1$
\cite{Ton,Tu1,Pee}, while in general case $w_{X}$ may vary with
time (models with quintessence) \cite{Pee,Liv}. Even if regarding
baryon component one can assume that it decreases as $a^{-3}$
(pressure of baryons may be neglected due to their relative small
amount in the universe), for dark matter (whose nature and
properties can be extracted only from its gravitational action on
ordinary matter) such a dependence on the scale factor may not
hold in the universe taken as a whole (in contrast to local
manifestations e.g. in large-scale structure formation, where
dependence $a^{-3}$ may survive). Since the contribution from all
baryons into the total energy density does not exceed 4 \%
\cite{Hag}, the evolution of the universe as a whole is determined
mainly by the properties of dark matter and dark energy. The
models of dark energy \cite{Tu1,Pee,Tu2} show explicitly unusual
behaviour of this component during the expansion of the universe.

In the present paper we notice that the quantum model of the
homogeneous and isotropic universe filled with primordial matter
in the form of the uniform scalar field proposed in Refs.
\cite{K,KK} allows to explain the observed coordinate distances to
SNe Ia and radio galaxies (RGs) in wide redshift range.

\section{Coordinate distance to source}
\label{Coor}

According to quantum model \cite{K,KK} the universe can be both in
quasistationary and continuum states. Quasistationary states are
the most interesting, since the universe in such states can be
characterized by the set of standard cosmological parameters
\cite{KK}. The wavefunction of quasistationary state as a function
of $a$ has a sharp peak and it is concentrated mainly in the
region limited by the barrier formed by the interaction between
the gravitational and scalar fields. It can be normalized (cf.
Ref. \cite{Foc}) and used in calculations of expectation values of
operators corresponding to observed parameters within the lifetime
of the universe.

If one considers the average value of the scale factor $\langle a
\rangle$ in the state with large quantum numbers, enumerating the
states of gravitational ($n$) and scalar ($s$) fields, as
determining the scale factor in classical approximation, then from
functional equation of quantum model for the wavefunction
\cite{K,KK} follows the Einstein-Friedmann equation in terms of
average values
\begin{equation}\label{1}
    \left(\frac{1}{\langle a \rangle}\,\frac{d \langle a
    \rangle}{dt}\right)^{2} = \langle \rho \rangle - \frac{1}{\langle a
    \rangle^{2}},
\end{equation}
where
\begin{equation}\label{2}
    \langle \rho \rangle = \gamma \, \frac{M}{\langle a \rangle^{3}} +
    \frac{E}{\langle a \rangle^{4}}
\end{equation}
is the mean total energy including the contribution from matter
and radiation in the universe in the states with $n \gg 1$ and $s
\gg 1$, and length and density are measured in units $l_{P} =
\sqrt{2G/(3\pi)}$ and $\rho_{P} = 3/(8 \pi G l_{P}^{2})$
respectively. The coefficient $\gamma = 193/12$ arises in
calculation of expectation value for the operator of energy
density of scalar field and takes into account its kinetic and
potential terms. The value $M = m \left (s + \frac{1}{2} \right
)$, where $m$ is a mass of elementary excitation of the scalar
field\footnote{As usual it is assumed that the scalar field
oscillates with a small amplitude around the equilibrium vacuum
value due to the quantum fluctuations.}, $s$ counts the number of
these excitations, can be treated as a quantity of matter/energy
in the universe. There is the following relation between the
parameters of the universe \cite{K,KK}
\begin{equation}\label{3}
    E = 4 \langle a \rangle \left [\langle a \rangle - M
    \right ].
\end{equation}
In matter dominated universe $M \gg E/(4 \langle a \rangle)$ and
from Eqs. (\ref{2}) and (\ref{3}) it follows that the quantity of
matter/energy $M$ and the mean energy density $\langle \rho
\rangle$ in the universe taken as a whole (i.e. in quantum states
which describe only homogenized properties of the universe)
satisfy the relations
\begin{equation}\label{4}
    M = \langle a \rangle, \qquad
    \langle \rho \rangle = \frac{\gamma }{\langle a \rangle^{2}}
\end{equation}
which agree with the data of observations in the present-day
universe. Substitution of Eq. (\ref{4}) into (\ref{1}) leads to
the density parameter $\Omega = 1.066$. It means that the universe
in highly excited states is spatially flat (to within about $7$
\%). This value of $\Omega$ agrees with existing astrophysical
data for the present-day universe \cite{Be,Kra,Ol}.

From (\ref{1}) and (\ref{4}) we find the Hubble constant $H =
(1/\langle a \rangle)\,d \langle a \rangle/dt$ as a function of
cosmological redshift $z = (a_{0}/\langle a \rangle) - 1$, where
$a_{0}$ is a scale factor at the present epoch,
\begin{equation}\label{5}
    H(z) = H_{0}\,(1 + z).
\end{equation}
Then the dimensionless coordinate distance $H_{0}\,r(z)$ to source
at redshift $z$, where $r(z) = (1 + z)^{-1}\,d_{L}$, $d_{L}$ is
the luminosity distance, (see Refs. \cite{Tu1,Vish,DD,We}) for a
flat universe obeys the logarithmic law
\begin{equation}\label{6}
    H_{0}\,r(z) = \ln (1 + z).
\end{equation}

\begin{figure}[ht]
\begin{center}
\includegraphics*{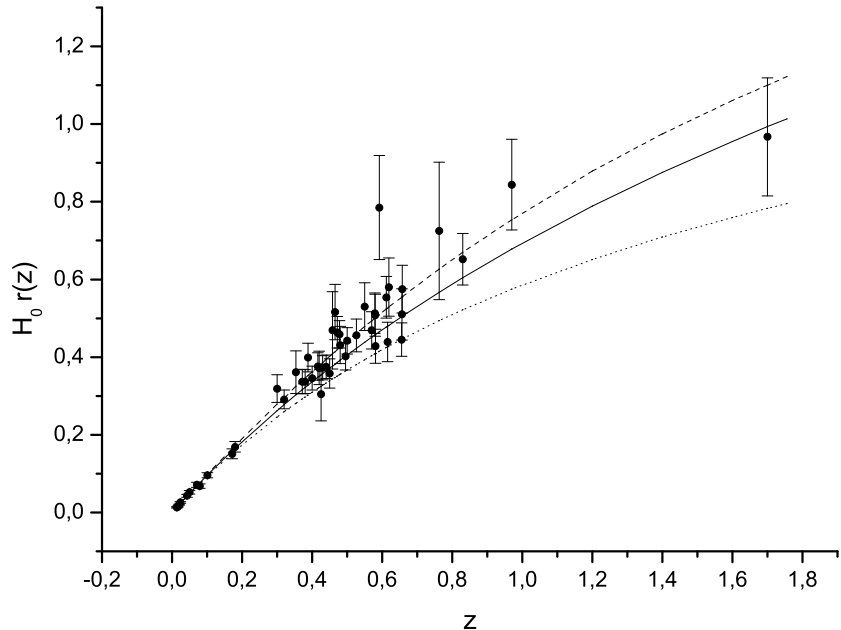}
\end{center}
\caption{Dimensionless coordinate distances $H_{0}\,r(z)$ to
supernovae at redshift $z$. The observed SNe Ia are shown as solid
circles. The model (\ref{4}) is drawn as a solid line. The
$\Lambda$CDM-model with $\Omega_{M} = 0.3$ and $\Omega_{X} = 0.7$
is represented as a dashed line. The model with $\Omega_{M} = 1$
is shown as a dotted line.} \label{fig:1}
\end{figure}

The dimensionless coordinate distances to the SNe Ia and RGs
obtained in Ref. \cite{DD} from the observational data (solid
circles and boxes) and our result (\ref{6}) (solid line) are shown
in Figs. 1 - 3. The $\Lambda$CDM-model with $\Omega_{M} = 0.3$
(matter component) and $\Omega_{X} = 0.7$ (dark energy in the form
of cosmological constant) and the model without dark energy
($\Omega_{M} = 1$) are drawn for comparison. In Fig. 2 the SNe Ia
data in the interval $z = 0.1 - 0.8$ are shown on a larger scale.
Among the supernovae shown in Figs. 1 and 2 there are the objects
with central values of coordinate distances which are better
described by the $\Lambda$CDM-model (e.g. 1994am at $z = 0.372$;
1997am at $z = 0.416$; 1995ay at $z = 0.480$; 1997cj at $z =
0.500$; 1997H at $z = 0.526$; 1997F at $z = 0.580$), the law
(\ref{6}) (e.g. 1995aw at $z = 0.400$; 1997ce at $z = 0.440$;
1995az at $z = 0.450$; 1996ci at $z = 0.495$; 1996cf at $z =
0.570$; 1996ck at $z = 0.656$) and the model with $\Omega_{M} = 1$
(1994G at $z = 0.425$; 1997aj at $z = 0.581$; 1995ax at $z =
0.615$; 1995at at $z = 0.655$). The RG data \cite{DD} demonstrate
the efficiency of the model (\ref{4}) as well (Fig. 3). The
quantum model predicts the coordinate distance to SN 1997ff at $z
\sim 1.7$ which is very close to the observed value (see Fig. 1).
In the range $z \leq 0.2$ three above mentioned models give in
fact the same result.

\begin{figure}[ht]
\begin{center}
\includegraphics*{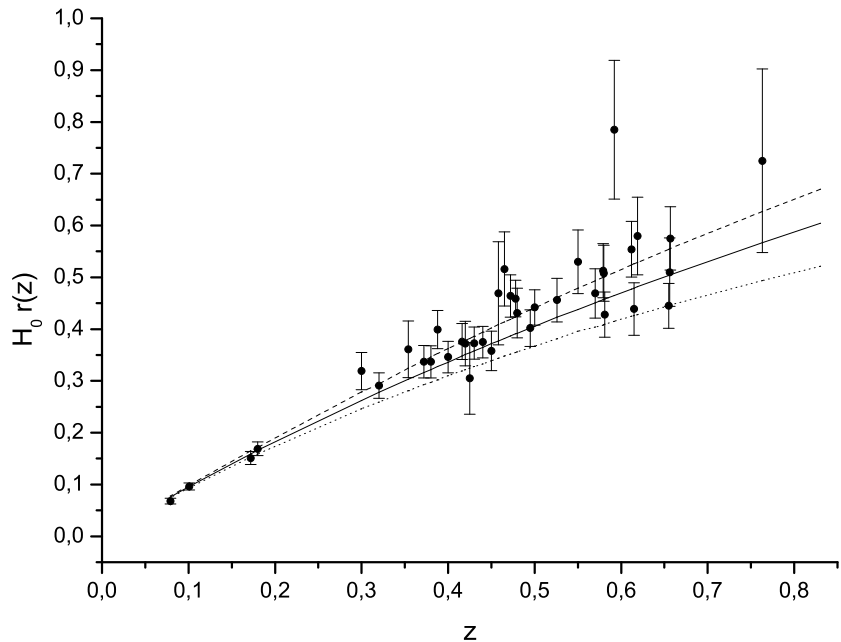}
\end{center}
\caption{The same as in Fig.1 in the interval $z = 0.1 - 0.8$ on a
larger scale.} \label{fig:2}
\end{figure}

The density $\langle \rho \rangle$ (\ref{2}) contains all possible
matter/energy components in the universe. Let us separate in
(\ref{2}) the baryon matter density equal to $\Omega_{B} \approx
0.04$ \cite{Kra,Ol} which makes a small contribution to the matter
density $\Omega_{M} \approx 0.3$ \cite{Ton}. If we assume that the
baryon density varies as $\rho_{B} \sim a^{-3}$, while the
remaining constituents of density effectively decrease as
$a^{-2}$, then the value $H_{0} r(z)$ calculated in such a model
will practically coincide with the coordinate distance shown in
Fig. 1 as a solid line.

\begin{figure}[ht]
\begin{center}
\includegraphics*{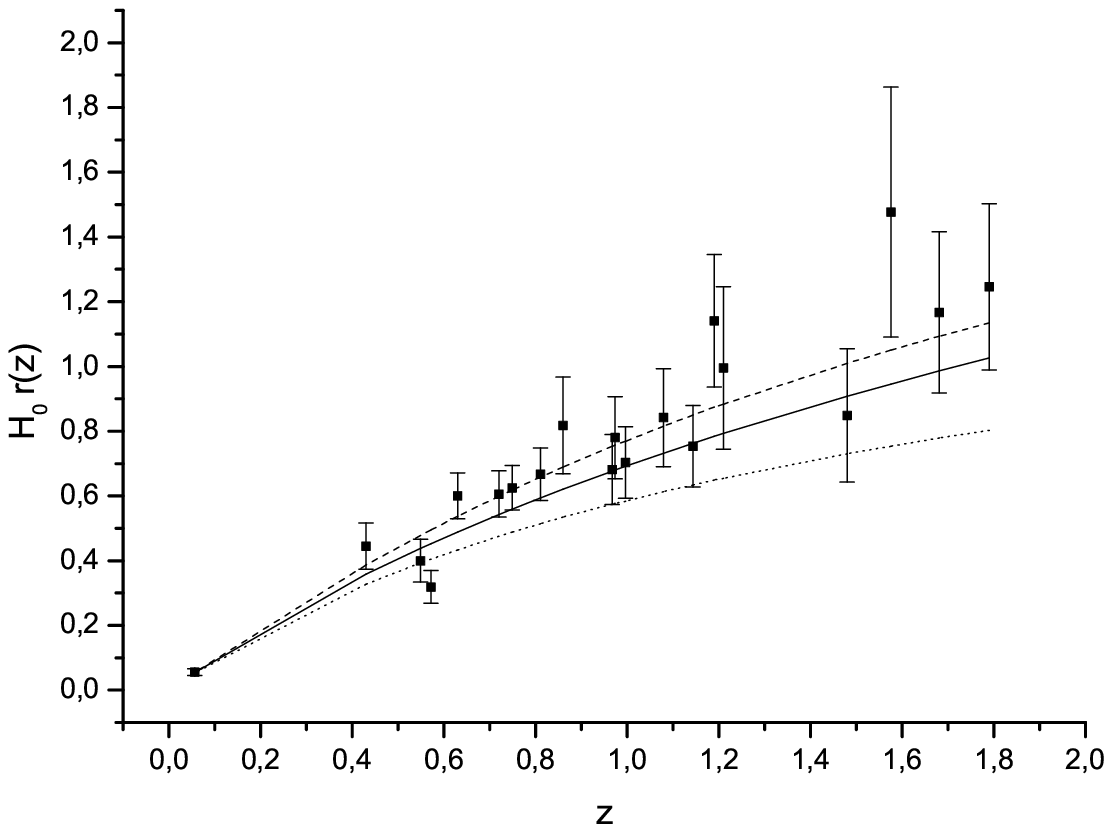}
\end{center}
\caption{Dimensionless coordinate distances $H_{0}\,r(z)$ to radio
galaxies at redshift $z$. Radio galaxies are shown as solid boxes.
The rest as in Fig.1.} \label{fig:3}
\end{figure}

In Ref. \cite{Vish} a conclusion is drawn that the model of dark
energy with $w_{X} = - \frac{1}{3}$ implying $\rho_{X} \sim
a^{-2}$ agrees with the recent CMB observations made by WMAP as
well as with the high redshift supernovae Ia data. Such a universe
is decelerating. In our quantum model the total dark matter/energy
in the states which describe only homogenized properties of the
universe varies effectively as $a^{-2}$. In terms of general
relativity it means that its negative pressure compensates for
action of gravitational attraction and the universe as a whole
expands at a steady speed.

\section{Quantum fluctuations of scale factor}
\label{Quan}

Deviations of $H_{0}\,r(z)$ from the law (\ref{6}) towards both
larger and smaller distances for some supernovae can be explained
by the local manifestations of quantum fluctuations of scale
factor about $\langle a \rangle$ which arose in the Planck epoch
($t \sim 1$) due to finite widths of quasistationary states. As it
is shown in Refs. \cite{K,KK} such fluctuations can cause the
formation of nonhomogeneities of matter density which have grown
with time into the observed large-scale structures in the form
superclusters and clusters of galaxies, galaxies themselves etc.
Let us consider the influence of mentioned fluctuations on visible
positions of supernovae.

The position of quasistationary state $E_{n}$ can be determined
only approximately, $E_{n} \rightarrow E_{n} + \delta E_{n}$,
where $|\delta E_{n}| \sim \Gamma_{n}$, $\Gamma_{n}$ is the width
of the state. The scale factor of the universe in the $n$-th state
can be found only with uncertainty,
\begin{equation}\label{7}
    \langle a \rangle \rightarrow \langle a \rangle + \delta a,
\end{equation}
where the deviation $\delta a \gtrless 0$ is determined by both
the value $\delta E_{n}$ and the time of its formation
\cite{K,KK}. Since $\Gamma_{n}$ is exponentially small for the
states $n \gg 1$, the fluctuations $\delta E_{n}$ in the early
universe are the main source for $\delta a$. The calculations
demonstrate that the lowest quasistationary state has the
parameters $E_{n=0} = 2.62$ and $\Gamma_{n=0} = 0.31$ (in
dimensionless units). The radius of curvature is $\langle a
\rangle_{n=0} \sim 1$, while the lifetime of such a universe is
$\tau \sim \Gamma_{n=0}^{-1} \sim 3$. Within the time interval
$\Delta t \leq 3$ the nonzero fluctuations of scale factor with
relative deviation equal e.g. to
\begin{eqnarray}
    \nonumber
    \left|\frac{\delta a}{\langle a \rangle}\right| & \lesssim & 0.022
    \quad
    \mbox{at} \ \ \Delta t=1, \\
    \label{8}
    \left|\frac{\delta a}{\langle a \rangle}\right| & \lesssim & 0.040
    \quad
    \mbox{at} \ \ \Delta t=2, \\
    \nonumber
    \left|\frac{\delta a}{\langle a \rangle}\right| & \lesssim & 0.077
    \quad
    \mbox{at} \ \ \Delta t=3
\end{eqnarray}
can be formed in the universe. Such fluctuations of the scale
factor cause in turn the fluctuations of energy density which can
result in formation of structures with corresponding linear
dimensions under the action of gravitational attraction. For
example, for the current value $\langle a \rangle \sim 10^{28}$ cm
the dimensions of large-scale fluctuations $\delta a \lesssim 70$
Mpc, $\delta a \lesssim 120$ Mpc, and $\delta a \lesssim 200$ Mpc
correspond to relative deviations (\ref{8}). On the order of
magnitude these values agree with the scale of superclusters of
galaxies. Let us note that the probability of the fluctuations
which can yield the scale of nonhomogeneities of the matter
density $\sim 200$ Mpc in visible part of the universe is small
because of large probability of tunneling of the universe into the
region of the continuous spectrum from the lowest state with $n =
0$ at $\Delta t=3$.

If one assumes that just the fluctuations $\delta a$ cause
deviations of positions of sources at high redshift from the law
(\ref{6}), then it is possible to estimate the values of relative
deviations $\delta a/\langle a \rangle$ from the observed values
$H_{0}\,r(z)$. The fluctuations of scale factor (\ref{7}) generate
the changes of coordinate distances,
\begin{equation}\label{9}
    H_{0}\,r(z) = \ln \left[\left(1 +
    \frac{\delta a}{\langle a \rangle} \right)^{-1} (1 + z)\right].
\end{equation}

\begin{table}[ht]
\caption{Relative deviations $\delta a/\langle a \rangle$ for some
type Ia supernovae}
\begin{center}
\begin{tabular}{ccccc} \hline
Group & SN Ia & $z$ & $H_{0}\,r(z)$ \cite{DD} & $\delta a/\langle
a \rangle$ \\ \hline 1 & 1994am & 0.372 & 0.337 & - 0.020 \\ &
1997cj & 0.500 & 0.442 & - 0.036 \\ & 1997H & 0.526 & 0.456 & -
0.033 \\ & 1997F & 0.580 & 0.508 & - 0.049 \\ & 1997R &
0.657 & 0.575 & - 0.068 \\ & 1997ck & 0.970 & 0.844 & - 0.153 \\
\hline 2 & 1997ff & 1.700 & 0.967 & + 0.027 \\ \hline 3 & 1994G &
0.425 & 0.305 & + 0.050 \\ & 1997aj & 0.581 & 0.428 & + 0.031 \\ &
1995ax & 0.615 & 0.439 & + 0.041 \\ & 1995at & 0.655 & 0.445 & +
0.061 \\ \hline 4 & 1996J & 0.300 & 0.319 & - 0.055 \\ & 1995ba &
0.388 & 0.399 & - 0.069 \\ & 1995ar & 0.465 & 0.516 & - 0.095 \\ &
1997K & 0.592 & 0.785 & - 0.274 \\ \hline
\end{tabular}
\end{center}
\end{table}

Relative deviations $\delta a/\langle a \rangle$ calculated from
observed coordinate distances (central values) for some type Ia
supernovae are shown in Table 1. The supernovae better described
by the $\Lambda$CDM-model are placed in Group 1. All shown
supernovae have negative relative deviations with absolute values
within the limits of (\ref{8}). The only exception is SN 1997ck at
$z = 0.970$. The supernova 1997ff at $z \sim 1.7$ which has the
coordinate distance close to the prediction of the quantum model
(\ref{6}) is located in Group 2. It has small positive relative
deviation within the bounds of (\ref{8}). The sources described by
the model with $\Omega_{M} = 1$ are shown in Group 3. They have
positive relative deviations and their values agree with (\ref{8})
as well. In Group 4 one can find four supernovae with coordinate
distances which are not described by any above mentioned model
even with the regard for observation errors. The values $\delta
a/\langle a \rangle$ for two of them (see Table 1) satisfy the
estimations (\ref{8}). For SNe 1997ck (Group 1) and 1995ar (Group
4) the values of relative deviations with regard for uncertainty
of measurements \cite{DD} (if one takes the lowest value) are
equal to $\delta a/\langle a \rangle = - 0.048$ and $\delta
a/\langle a \rangle = - 0.061$ respectively. The supernova 1997K
at $z = 0.592$ is characterized by too sharp negative relative
deviation even in comparison with the largest possible
fluctuations of the scale factor. If one takes into account
uncertainty of measurements and accept for distance the value
$H_{0} r(z) = 0.651$ \cite{DD} then the corresponding deviation
$\delta a/\langle a \rangle = - 0.17$ will still exceed the
relative deviation at $\Delta t = 3$. The same analysis one can
make for RGs as well.

Thus the observed faintness of some SNe Ia can in principle be
explained by the logarithmic-law dependence of coordinate distance
on redshift in generalized form (\ref{9}) which takes into account
the fluctuations of scale factor about its average value. These
fluctuations can arise in the early universe and grow with time
into observed deviations of the coordinate distances of separate
SNe Ia at the high redshift. They produce accelerating or
decelerating expansions of space subdomains containing such
sources whereas the universe as a whole expands at a steady rate.

\section{Concluding remarks}
\label{Con}

In addition to the prediction about the steady-speed expansion of
the universe as a whole (at the same time the accelerating or
decelerating motions of its subdomains remain possible on a
cosmological scale as it is shown in Sect. \ref{Quan}) the quantum
model allows an increase of quantity of matter/energy in matter
dominated universe according to (\ref{3}). If the mass $m$ of
elementary excitations of the scaler field remains unchanged
during the expansion of the universe, then the increase of $M$ can
occur due to increase in number $s$ of these excitations. But the
increase in $s$ does not mean that a quantity of observed matter
in some chosen volume of the universe increases. According to the
model proposed in Refs. \cite{KK2,KK3} the observed ``real''
matter (both luminous and dark) is created as a result of the
decay of excitations of the scalar field (under the action of
gravitational forces) into baryons, leptons and dark matter. The
undecayed part of them forms what can be called a dark energy.
Such a decay scheme leads to realistic estimates of the percentage
of baryons, dark matter and energy in the universe with $\langle a
\rangle \gg 1$ and $M \gg 1$. Despite the fact that the quantity
of matter/energy can increase, the mean total energy density
decreases and during the expansion of the universe mainly the
number of elementary excitations of the scalar field increases.
Their decay probability is very small, so that basically only the
dark energy is created. These circumstances can explain the
absence of observed events of creation of a new baryonic matter on
a cosmologically significant scale.

The proposed approach to the explanation of observed dimming of
some SNe Ia may provoke objections in connection with the problem
of large-scale structure formation in the universe, since the
energy density $\langle \rho \rangle$ in the form (\ref{4}) cannot
ensure an existence of a growing mode of the density contrast
$\delta \langle \rho \rangle/\langle \rho \rangle$ (see e.g. Refs.
\cite{Ol,We,Pee2}). As we have already mentioned above in Sect.
\ref{Coor} the density $\langle \rho \rangle$ (\ref{4}) describes
only homogenized properties of the universe as a whole. It cannot
be used in calculations of fluctuations of energy density about
the mean value $\langle \rho \rangle$. Under the study of
large-scale structure formation one should proceed from the more
general expression for the energy density (\ref{2}). Defining
concretely the contents of matter/energy $M$, as for instance in
the model of creation of matter mentioned above, one can make
calculations of density contrast as a function of redshift. The
problem of large-scale structure formation is one of the main
problems of cosmology (see e.g. Refs. \cite{Ol,Dol}). It goes
beyond the tasks of this paper and requires a special
investigation. The ways of its solution in the quantum model are
roughly outlined in Ref. \cite{KK}.


\begin{thebibliography}{00}

\bibitem{Rie} A.G. Riess et al., Astron J. \textbf{116}, 1009
(1998), astro-ph/9805201.

\bibitem{Per} S. Perlmutter et al., Astrophys. J. \textbf{517},
565 (1999), astro-ph/9812133; Int. J. Mod. Phys. \textbf{A15}, S1,
715 (2000).

\bibitem{Ton} J.L. Tonry et al., Astrophys. J. \textbf{594}, 1 (2003),
astro-ph/0305008.

\bibitem{Tu1} M.S. Turner, in Type Ia Supernovae: Theory and Cosmology, eds.
J.C. Nieweyer and J.B. Truran (Cambridge University Press, 2000),
astro-ph/9904049; M.S. Turner and A.G. Riess, astro-ph/0106051.

\bibitem{Pee} P.J.E. Peebles and B. Ratra, Rev. Mod. Phys. \textbf{75}, 599
(2003), astro-ph/0207347.

\bibitem{Liv} M. Livio, astro-ph/0303500.

\bibitem{Vish} R.G. Vishwakarma, Mon. Not. Roy. Astron. Soc. \textbf{345}, 545
(2003), astro-ph/0302357.

\bibitem{Tu2} M.S. Turner, Astrophys. J. \textbf{576},
L101 (2002), astro-ph/0106035; Int. J. Mod. Phys. \textbf{A17},
3446 (2002), astro-ph/0202007; astro-ph/0207297.

\bibitem{Ost} J.P. Ostriker and P.J. Steinhardt, Nature
\textbf{377}, 600 (1995), astro-ph/9505066; N.A. Bancall, J.P.
Ostriker, S. Perlmutter, and P.J. Steinhardt, Science
\textbf{284}, 1481 (1999), astro-ph/9906463.

\bibitem{Hag} K. Hagiwara et al., Phys. Rev. \textbf{D66},
010001-1 (2002).

\bibitem{K} V.V. Kuzmichev, Ukr. J. Phys. \textbf{43}, 896 (1998);
Phys. At. Nucl. \textbf{62}, 708 (1999), gr-qc/0002029; Phys. At.
Nucl. \textbf{62}, 1524 (1999), gr-qc/0002030.

\bibitem{KK} V.E. Kuzmichev and V.V. Kuzmichev, Eur. Phys. J.
\textbf{C23}, 337 (2002), astro-ph/0111438.

\bibitem{Foc} V.A. Fock, Nachala kvantovoi mekhaniki (Foundation
of Quantum Mechanics) (Nauka, Moscow, 1976).

\bibitem{Be} P. deBernardis et al., Nature \textbf{404}, 955
(2000); C.B. Netterfield et al.,  Astrophys. J. \textbf{571}, 604
(2002), astro-ph/0104460; C. Pryke et al., Astrophys. J.
\textbf{568}, 46 (2002), astro-ph/0104490; J.L. Sievers et al.,
Astrophys. J. \textbf{591}, 599 (2003), astro-ph/0205387.

\bibitem{Kra} L.M. Krauss, astro-ph/0301012.

\bibitem{Ol} K.A. Olive, astro-ph/0301505.

\bibitem{DD} R.A. Daly and S.G. Djorgovski, Astrophys. J.
\textbf{597}, 9 (2003), astro-ph/0305197.

\bibitem{We} S. Weinberg, Gravitation and Cosmology (Wiley, New
York, 1972).

\bibitem{KK2} V.E. Kuzmichev and V.V. Kuzmichev, Ukr. J. Phys.
\textbf{48}, 801 (2003), astro-ph/0301017; astro-ph/0302173.

\bibitem{KK3} V.E. Kuzmichev and V.V. Kuzmichev, in Selected
Topics in Theoretical Physics and Astrophysics, eds. A.K.
Motovilov, F.M. Pen'kov (JINR, Dubna, 2003) 136.

\bibitem{Pee2} P.J.E. Peebles, The Large-Scale Structure of the
Universe (Princeton Univ. Press, Princeton, 1980).

\bibitem{Dol} A.D. Dolgov, hep-th/0306200.

\end{thebibliography}
\end{document}